\documentstyle[preprint,eqsecnum,aps,epsf,floats,tighten]{revtex} 

\begin{document}

\preprint{\tighten \vbox{\hbox{hep-ph/9903510} 
		\hbox{} \hbox{} }}

\title{Long distance effects on the $B\to X_s\gamma$ photon energy spectrum}

\author{Changhao Jin}

\address{School of Physics, University of Melbourne\\
Parkville, Victoria 3052, Australia}

\maketitle

{\tighten
\begin{abstract}%
We compute long distance effects on the photon energy spectrum in inclusive 
radiative decays of $B$ mesons using light-cone expansion and heavy quark
effective theory. We show that for sufficiently high photon energy the leading
nonperturbative QCD contribution is attributed to the distribution function. 
The distribution function is found to be universal in the sense that the same 
distribution function also encodes the leading nonperturbative contribution to
inclusive semileptonic decays of $B$ mesons at large momentum transfer. Some 
basic properties of the distribution function are deduced in QCD. Ways of 
extracting the distribution function directly from experiment and their 
implications are discussed. The theoretically clean methods for the 
determination of $|V_{ts}|$ are described. 
\end{abstract}
}

\newpage

\section{Introduction}
The study of the inclusive radiative decay $B\to X_s\gamma$, where $X_s$ is 
any possible final state of total strangeness $-1$, is a broad subject with 
many areas of investigation.
It is a flavour changing neutral current process, which is forbidden at tree
level, only proceeding through what is called a electroweak penguin diagram 
\cite{vainsh,ellis,bander,guber,inami,deshp,campb,nazer} 
in the Standard Model. This is a one-loop graph inducing $b\to s +$photon 
with a $W$ boson and a quark, predominantly the top quark, in the loop.  
Measurements of this rare
process provide one of the most stringent experimental tests of the Standard 
Model at one-loop level. It can be used to determine the 
fundamental Cabibbo-Kobayashi-Maskawa (CKM) mixing
matrix element $|V_{ts}|$ and test its unitarity. It may offer new insight
into the nature of confinement. 

Meanwhile, because the decay is highly suppressed in the Standard Model, 
it could be particularly sensitive to new physics beyond
the Standard Model. Measurements of $B\to X_s\gamma$ decays 
would impose constraints on new physics models.  In particular,
new contributions from nonstandard particles replacing the standard model 
particles in the loop can be detected prior to a direct production of such new
particles at much higher energies. 
A clear deviation from standard model expectations would signal new physics
from supersymmetry, charged Higgs scalars, anomalous $WW\gamma$ couplings, etc.

Observation of the inclusive radiative decay $B\to X_s\gamma$ was reported 
in 1995 by the CLEO Collaboration, obtaining the first measurement of 
the branching ratio of $(2.32\pm 0.57\pm 0.35)\times 10^{-4}$ \cite{cleo}. 
The ALEPH Collaboration has reported a
measurement of the corresponding branching ratio of beauty hadrons produced
at the Z resonance, obtaining the branching ratio of 
$(3.11\pm 0.80\pm 0.72)\times 10^{-4}$ \cite{aleph}.
Recently, the CLEO Collaboration 
has improved their measurement using $60\%$ additional
data and improved analysis techniques, yielding a new preliminary result
of the branching ratio of 
$(3.15\pm 0.35\pm 0.32\pm 0.26)\times 10^{-4}$ \cite{newcleo}.  
Copious data samples from the rare decays $B\to X_s\gamma$ will soon be
collected at $B$ factories. The accuracy of the $B\to X_s\gamma$ measurement
will be significantly improved shortly.

Significant progress has been made in the theoretical prediction of the 
branching ratio for $B\to X_s\gamma$ in the Standard Model. 
The next-to-leading order (NLO) calculation of the
perturbative QCD corrections has been completed recently, combining the
calculations of the matching conditions of the Wilson 
coefficients \cite{adel,gh,bkp,ciuch}, 
the matrix elements \cite{ali1,pott,greub}, and
the anomalous dimensions \cite{chety}.
This achievement leads to a substantial reduction of the large 
renormalization scale dependence of the leading order 
result \cite{bertol,plo,grigj,grin,chiuch1,cella,misiak,ali2,bmmp}. 
The leading power corrections in $1/m_b^2$ \cite{bbsuv,mb2} and 
$1/m_c^2$ \cite{volo,khod,ligeti,grant,buch}
have been calculated in the context of the heavy quark expansion.
In addition, the leading electroweak radiative corrections
have recently been investigated \cite{czarn,kagan}. 

It is, however, not sufficient to have a reliable calculation of the total
decay rate. Experimentally,
in order to suppress backgrounds from other $B$-decay processes, only
events in the high energy region of the photon energy spectrum in $B\to X_s
\gamma$ decays have been used to measure the branching ratio. An accurate and
reliable 
theoretical description of the photon energy spectrum is essential in order
to perform a fit to the experimental data and extrapolate to the total decay 
rate. The study of the photon energy spectrum is also quite interesting in its
own right, since it would probe the decay dynamics in more detail than the 
total rate.  Strong interaction effects
can significantly modify the photon energy spectrum. The following fact well 
illustrates the situation.  It is well known that the quark-level and virtual 
gluon exchange processes $b\to s\gamma$ generate a trivial photon energy 
spectrum --- a discrete line at $E_\gamma= m_b/2$ in the $b$-quark rest 
frame.  It is two distinct effects, gluon bremsstrahlung and hadronic bound 
state effects, that spread out the spectrum. Gluon bremsstrahlung results in
a long tail in the photon spectrum below $m_b/2$. Bound state effects leads to
the extension of 
phase space from the parton level to the hadron level, also stretches 
the spectrum downward below $m_b/2$ and is solely responsible for populating 
the spectrum upward in the gap between the parton level endpoint 
$E_\gamma= m_b/2$ and the hadron level endpoint $E_\gamma= M_B/2$.

The perturbative QCD corrections to the photon energy spectrum are known at
order $\alpha_s$ \cite{greub} and the order $\alpha_s^2\beta_0$ contribution 
is computed \cite{wise} very recently in the perturbation expansion of the 
matrix elements. The relevant Wilson coefficients in the effective weak 
Hamiltonian are known to next-to-leading logarithmic accuracy \cite{chety}
in renormalization group improved perturbative QCD. 
Currently nonperturbative QCD effects comprise the potentially most 
serious source of theoretical error in the photon energy spectrum. 
Long distance QCD effects are relevant for improving the accuracy of an 
analysis and understanding theoretical uncertainties in 
the Standard Model, as well as for searching for new physics by the 
confrontation of the standard model predictions with the data. 
In view of this, it is important to address the issue as rigorously and
completely as possible.  The Fermi motion of the $b$ quark in the
$B$ meson has been taken into account in \cite{ali1} by using
the phenomenological model by Altarelli et al. \cite{accmm}. 
A more fundamental treatment of bound state effects based on the heavy quark 
expansion has been developed by resumming an infinite set of leading-twist 
operators into a shape function \cite{neubert,neubert1,shifman}. 
Recently, an analysis of long distance effects on the
photon spectrum has been performed \cite{kagan} along these lines, 
including for the first time the full NLO perturbative QCD corrections. 

In this paper we will compute the long distance QCD contributions to the
photon energy spectrum using light-cone expansion and heavy quark
effective theory (HQET). 
These ideas and techniques have originally been exploited in
the theoretical description of inclusive semileptonic decays of beauty 
hadrons \cite{jp}. This approach has been related \cite{vub} to the 
above-mentioned similar approach based on the resummation of the heavy quark 
expansion \cite{neubert,neubert1,shifman}. In this paper we extend our 
previous analyses to inclusive radiative decays of $B$ mesons. We strive to
present our approach in a more detailed and complete form.

The fact that beauty hadrons are the heaviest hadrons actually formed
since the top quark is too heavy to build hadrons 
confers a special role to beauty hadrons. The beauty hadron decays involve two
large scales: the heavy beauty hadron mass at the hadron level and
the heavy $b$ quark mass at the parton level, which are much greater than 
the energy scale $\Lambda_{\rm QCD}$ which characterizes the strong 
interactions. The two large scales give rise to two techniques for dealing
with nonperturbative QCD. Because of the heaviness of the decaying 
hadron, the decay dynamics may be dominated by the space-time separations
in the neighborhood of the light cone. 
The light-cone expansion provides a formal and powerful way of organizing the 
nonperturbative QCD effects and singling out the leading term. On the other
hand, the heavy quark mass provides a large limit to construct an effective
theory describing the heavy quark interacting with the gluons in the heavy
hadron. This so-called heavy quark effective theory has new (approximate)
symmetries that were not manifested in the full QCD Lagrangian and sets 
another 
framework for organizing and parametrizing nonperturbative effects, which
relates various phenomena (e.g., the hadron spectroscopy and weak decays of
hadrons containing a single heavy quark) to a common set of parameters, so
that it has great predictive power.  It is certainly useful to calculate long 
distance
QCD effects on the photon energy spectrum in $B\to X_s\gamma$ decays using
light-cone expansion and heavy quark effective theory. 

The paper is organized as follows.  In Sec.~II, we generalize the results of 
Ref.~\cite{jp} and construct the light-cone expansion for the photon energy 
spectrum in $B\to X_s\gamma$ decays.  We show that for sufficiently high 
photon energy, the leading nonperturbative QCD contribution
is attributed to a distribution function. We define a new, gauge-invariant 
distribution function. The connection between the 
nonperturbative QCD effects in radiative and semileptonic inclusive $B$ 
decays is subsequently established in Sec.~III.  For the sake of completeness,
we study the
properties of the new distribution function in Sec.~IV. We discuss the direct 
extraction of the distribution function from experiment and precise 
determinations of the CKM matrix elements in Sec.~V. We hope to motivate the
measurement of the distribution function by pointing out its importance in
inclusive $B$ decays and the ways to measure it. Section VI briefly summarizes
the main results.

\section{Light-Cone Expansion for $B\to X_{\lowercase{s}}\gamma$ Decays}
We shall study inclusive radiative decays $B(P)\to X_s (p_X)
\gamma (p_\gamma)$.
A suitable framework to do that is an effective low-energy theory,
obtained by integrating out the heavy particles which in the Standard Model
are the top quark and the W boson \cite{grin}.  
Since we shall concentrate on long distance effects, it suffices to work at 
lowest order in electroweak interactions ignoring for 
the moment perturbative QCD and electroweak radiative corrections.
To lowest order $B\to X_s\gamma$ decays are governed by the effective 
weak Hamiltonian involving a magnetic dipole operator: 
\begin{equation}
{\cal H}_{\rm eff}= \kappa\bar s\sigma^{\mu\nu}RbF_{\mu\nu} ,
\label{eq:heff}
\end{equation}
where the coupling constant
\begin{equation}
\kappa= -\frac{G_Fem_b}{4\sqrt{2}\pi^2}V_{tb}V^\ast_{ts}C^{(0)}_7(M_W)
\label{eq:kappa} 
\end{equation}
gauges the strength of the $b\to s\gamma$ transition, $R= (1+\gamma_5)/2$ 
is the right-handed projection operator,
and $F_{\mu\nu}$ is the electromagnetic field strength tensor. In 
Eq.~(\ref{eq:kappa}),
$V_{tb}$ and $V_{ts}$ are the CKM matrix elements,
the function $C^{(0)}_7 (M_W)$ depends on the masses of the internal top 
quark and W boson, and takes the form \cite{inami}
\begin{equation}
C^{(0)}_7(M_W)=\frac{3x_t^3-2x_t^2}{4(x_t-1)^4}{\rm ln} x_t-
\frac{8x_t^3+5x_t^2-7x_t}{24(x_t-1)^3} 
\label{eq:c7}
\end{equation}
with $x_t= m_t^2/M_W^2$. We use $m_b$ to
denote the $b$ quark mass and $M_B$ to denote the $B$ meson mass.
It can be shown that contributions of the strange quark mass to the decay
rate are suppressed by the very small factor $m_s^2/m_b^2$. Hence we neglect 
the strange quark mass unless noted otherwise.

The decay rate for $B\to X_s\gamma$ is given by
\begin{equation}
d\Gamma= \frac{1}{2E_B}\frac{d^3{\bf p}_\gamma}{(2\pi)^3 2E_\gamma}
\sum_{X_s,\epsilon}(2\pi)^4\delta^4(P-p_\gamma-p_X)
|\langle X_s(p_X)\gamma (p_\gamma, \epsilon)|{\cal H}_{\rm eff}|
B(P)\rangle|^2 ,
\label{eq:rate1}
\end{equation}
where we sum over all possible final states of total strangeness $-1$ as
well as the two transverse polarizations of the photon.
We adopt the standard covariant normalization 
$\langle B|B\rangle=2E_B(2\pi)^3\delta^3({\bf 0})$ for the $B$ meson state. 
The decay rate can be expressed in terms of a 
current commutator taken between the $B$ meson states, which incorporates all
nonperturbative QCD physics of the weak radiative $B$-meson decay. We find
\begin{equation}
d\Gamma= \frac{|\kappa|^2}{2E_B}\frac{d^3{\bf p}_\gamma}{(2\pi)^3 2E_\gamma}
\int d^4 y\, e^{ip_\gamma \cdot y}
\langle B\left|[J^{\dagger}_{\mu}(y), J^\mu (0)]\right|B\rangle ,
\label{eq:rate2}
\end{equation}
with the generalized current $J_{\mu}(y) = \bar{b}(y)[\gamma_{\mu}, 
p_{\gamma}\!\!\!\!\!/ \,\,] L s(y)$, where $L= (1-\gamma_5)/2$ is the 
left-handed projection operator.

The commutator in Eq.~(\ref{eq:rate2}) 
has to vanish for space-like separations $y^2<0$ due to causality. 
Moreover, the behavior of the integral in 
Eq.~(\ref{eq:rate2}) is determined by the integrand in domains with less
rapid oscillations, i.e., $|p_\gamma\cdot y|\sim 1$, which implies 
$y^2\lesssim 1/E^2_\gamma$. Combining these results we find that the dominant
contribution to the integral in Eq.~(\ref{eq:rate2}) results from the
space-time region $0\leq y^2\lesssim 1/E^2_\gamma$. This implies that for
sufficiently high photon energy, $E_\gamma\gg\Lambda_{\rm QCD}$, the
space-time separations in the neighborhood of the light-cone $y^2=0$ dominate
the decay dynamics. It is then appropriate
to construct a light-cone expansion to calculate the photon energy spectrum in
the high energy region.  

The decay produced strange quark propagating in a background
gluon field obeys the anticommutation relation
\begin{equation}
\{s(x), \bar s(y)\}= (i\partial_x\!\!\!\!\!/ +m_s) i\Delta_s(x-y) U(x, y),
\label{eq:strange}
\end{equation}
with the Wilson link 
\begin{equation}
U(x, y)= {\cal P}{\rm exp}[ig_s\int_y^x dz^{\mu}A_{\mu}(z)] 
\end{equation}
between the quark fields at $y$ and $x$,
where ${\cal P}$ denotes path ordering. Here the strange quark mass $m_s\neq 0$
is retained. The background gluon field $A^\mu$ is that
generated by the remnants of the decaying $B$ meson.
$\Delta_q(y)$ is the Pauli-Jordan function of the form \cite{bogol}
\begin{eqnarray}
\Delta_q (y)&=& -\frac{i}{(2\pi)^3}\int d^4k\, e^{-ik\cdot y}\varepsilon (k^0)
\delta (k^2-m_q^2)  \nonumber \\
&=& -\frac{1}{2\pi}\varepsilon (y^0)\delta (y^2)+\frac{m_q}{4\pi\sqrt{y^2}}
\varepsilon (y^0)\theta (y^2)J_1 (m_q\sqrt{y^2}) .
\label{eq:pjfun}
\end{eqnarray}
Here $J_1 (z)$ is the Bessel function of order 1.  From Eq.~(\ref{eq:pjfun}) 
we see that
$\Delta_q (y)$ is singular on the light cone and would also select out 
light-cone contributions were it not for high frequency variations in the
phase.  Using the anticommutation relation (\ref{eq:strange}), we find
\begin{equation}
[J^{\dagger}_\mu (y), J^\mu (0)]= 16 p_{\gamma_{\alpha}}p_{\gamma_{\beta}}
[i\partial^\alpha i\Delta_s (y)]\bar b(0)\gamma^\beta RU(0, y)b(y) .
\label{eq:current}
\end{equation}

Substituting Eq.~(\ref{eq:current}) in Eq.~(\ref{eq:rate2}), it follows 
that
\begin{equation}
d\Gamma= \frac{|\kappa|^2}{2E_B}\frac{d^3{\bf p}_\gamma}{(2\pi)^3 2E_\gamma}
8p_{\gamma_{\alpha}}p_{\gamma_{\beta}}\int d^4 y\, e^{ip_\gamma \cdot y}
[i\partial^\alpha i\Delta_s (y)]\langle B|\bar b(0)\gamma^\beta
U(0, y)b(y)|B\rangle .
\label{eq:rate3}
\end{equation}
Note that the matrix element of $\bar b(0)\gamma^\beta\gamma_5 U(0, y)b(y)$ 
between the $B$ meson states vanishes by parity invariance in the
strong interactions.

Now let us consider the matrix element of the bilocal operator in 
Eq.~(\ref{eq:rate3}). The general tensor decomposition of it reads
\begin{equation}
\langle B|\bar b(0)\gamma^\beta U(0, y)b(y)|B\rangle 
= 2[P^\beta F(y^2, y\cdot P)+y^\beta G(y^2, y\cdot P)] ,
\label{eq:decomp}
\end{equation}
where $F(y^2, y\cdot P)$ and $G(y^2, y\cdot P)$ are in general functions of 
the two independent Lorentz scalars, $y^2$ and $y\cdot P$.
Contracting both sides of Eq.~(\ref{eq:decomp}) with $y_\beta$ gives
\begin{equation}
\langle B|\bar b(0)y\!\!\!/ U(0, y)b(y)|B\rangle 
= 2[y\cdot P F(y^2, y\cdot P)+y^2 G(y^2, y\cdot P)] .
\label{eq:contract}
\end{equation}
In the domain of interest where $|p_\gamma\cdot y|\sim 1$ and 
$0\leq y^2\lesssim 1/E^2_\gamma$, 
we infer from Eq.~(\ref{eq:contract}) that the contribution of 
$G(y^2, y\cdot P)$ is suppressed by a factor $1/E^2_\gamma$ for high photon
energy considered here, compared to that of $F(y^2, y\cdot P)$. 
In addition, we can make a light-cone expansion for $F(y^2, y\cdot P)$,
which is justified for high photon energy as discussed above,
\begin{equation}
F(y^2, y\cdot P)= \sum_{n=0}^\infty \frac{(y^2)^n}{n!}
\Bigg [\frac{d^n F(y^2, y\cdot P)}{dy^{2n}}\Bigg ]_{y^2=0} .
\label{eq:expan}
\end{equation}
The $n$th term in the light-cone expansion is suppressed by 
$1/E^{2n}_\gamma$. Therefore, in the leading-twist approximation we then have
\begin{equation}
\langle B|\bar b(0)\gamma^\beta U(0, y)b(y)|B\rangle 
= 2 P^\beta F(y^2=0, y\cdot P) .
\label{eq:leading}
\end{equation}
The contributions from $F(y^2\neq 0, y\cdot P)$ and $G(y^2, y\cdot P)$ 
represent
higher twist effects. They become important only at the subleading level, 
a question that will not be dwelt upon here.

On the light cone $F(y^2=0, y\cdot P)$ can be projected out of the general 
decomposition (\ref{eq:contract}):
\begin{equation}
F(y^2=0, y\cdot P)= \frac{1}{2y\cdot P}\langle B|\bar b(0)y\!\!\!/
U(0, y)b(y)|B\rangle |_{y^2=0} .
\label{eq:proj}
\end{equation}
The Fourier transform of $F(y^2=0, y\cdot P)$ defines the distribution
function
\begin{eqnarray}
f(\xi)&=& \frac{1}{2\pi}\int d(y\cdot P)\, e^{i\xi y\cdot P}
F(y^2=0, y\cdot P) \nonumber \\
&=& \frac{1}{4\pi}\int\frac{d(y\cdot P)}{y\cdot P}\, e^{i\xi y\cdot P}
\langle B|\bar b(0)y\!\!\!/ U(0, y)b(y)|B\rangle |_{y^2=0} .
\label{eq:def}
\end{eqnarray} 
The Wilson link is associated with the background gluon field, which ensures 
gauge invariance of the distribution function. The leading nonperturbative QCD 
contribution contained in $F(y^2=0, y\cdot P)$ is equivalently translated into 
the distribution function.

Using the inverse Fourier transform
\begin{equation}
F(y^2=0, y\cdot P)= \int d\xi\, e^{-i\xi y\cdot P}f(\xi) \, ,
\label{eq:inverse}
\end{equation}
substituting Eqs.~(\ref{eq:pjfun}) and (\ref{eq:leading}) in
Eq.~(\ref{eq:rate3}), and then carrying out the phase space integration,
we arrive at the photon energy spectrum in the $B$ rest frame
\begin{equation}
\frac{d\Gamma(B\to X_s\gamma)}{dE_\gamma}= \frac{G_F^2\alpha m_b^2}{2\pi^4 M_B}
|V_{tb}V_{ts}^\ast|^2 |C^{(0)}_7(M_W)|^2 
E_\gamma^3f\Bigg (\frac{2E_\gamma}{M_B}\Bigg ).
\label{eq:spectrum}
\end{equation}
Equation (\ref{eq:spectrum})
holds in the leading twist approximation, which is expected to be a
good approximation for the energy of the emitted photon of around 1 
GeV and above. The precise shape of the spectrum near the lower endpoint 
$E_\gamma= 0$ is
not available to us, where the light-cone expansion is not applicable and
higher twist terms give contributions of the same order as the leading twist
term. However, given the experimental \cite{cleo,aleph,newcleo} and 
theoretical \cite{ali1,shifman,kagan} indications that the spectrum for 
$E_\gamma < 1$ 
GeV appears to be vanishingly small, a more accurate account of the rather 
small overall nonperturbative contributions in this region seems numerically 
unimportant.
This is not unexpected since the photon energy spectrum stemming from the 
quark-level and virtual gluon exchange processes would only concentrate at
$E_\gamma= m_b/2\approx 2.45$ GeV and gluon bremsstrahlung and hadronic bound
state effects smear the spectrum about this point, but most of the decay rate
remains at large values of $E_\gamma$. 
For practical purposes, this approach would nonetheless provide a 
realistic description of the photon spectrum over the full kinematic range 
$0\leq E_\gamma\leq M_B/2= 2.64$ GeV. Of course, one should keep in mind that
for the photon spectrum near the lower endpoint one has nothing from the
approach to be verified or falsified.

Finally, let us discuss the relation of our approach to the approach advocated
in \cite{neubert,neubert1,shifman} based on the resummation of the heavy quark
expansion. By assuming $E_\gamma= m_b/2$ for the factor 
$E^3_\gamma$ in Eq.~(\ref{eq:spectrum}) we can reproduce their formulas for 
the $B\to X_s\gamma$
photon energy spectrum obtained in \cite{neubert1,shifman}.
In that case, the photon energy is 
fixed to the value in the free quark limit, instead of varying in its entire 
kinematic range from $0$ to $M_B/2$. 

\section{Universality}
In this section we compare the inclusive radiative decay with the inclusive 
semileptonic decays of $B$ mesons $B\to X_f\ell\nu$ $(f=u, c)$ in order to 
decipher the universal,
process-independent structure in the relevant realm of hadron physics.

Inclusive semileptonic decays of $B$ mesons at large momentum transfer carried
by the $W$ boson are also governed by the light-cone dynamics \cite{jp}.  
The nonperturbative QCD effects on the decays reside in the hadronic tensor
\begin{equation}
W_{\mu\nu} = -\frac{1}{2\pi} \int d^4y\, e^{iq\cdot y}\langle B|[j_{\mu}(y),
  j^{\dagger}_{\nu}(0)]|B\rangle ,
\label{eq:tensor}
\end{equation}
where the charged weak current 
$j_\mu(y)=\bar f(y)\gamma_\mu(1-\gamma_5)b(y)$.
Calculating the charged weak current commutator in the same way as in the
last section yields 
\begin{equation}
\langle B| \left[ j_{\mu}(y),j_{\nu}^{\dagger}(0) \right] |B\rangle
 = 2(S_{\mu\alpha\nu\beta} -i\varepsilon_{\mu\alpha\nu\beta})
  \left[ \partial^{\alpha}\Delta_f (y) \right] \langle B|\bar{b}(0)
    \gamma^{\beta}U(0, y)b(y)|B\rangle\, ,
\label{eq:domin3}
\end{equation}
where $S_{\mu\alpha\nu\beta} = g_{\mu\alpha}g_{\nu\beta} + g_{\mu\beta}
 g_{\nu\alpha} - g_{\mu\nu}g_{\alpha\beta}$. 
We observe here the very same matrix element, 
$\langle B|\bar b(0)\gamma^\beta U(0, y)b(y)|B\rangle$, entering the 
decay rate for the inclusive semileptonic decays of $B$ mesons.

The twist expansion of the bilocal operator matrix element then leads
to the expression of the differential decay rate in the $B$ rest frame in 
terms of the distribution function $f(\xi)$ defined in Eq.~(\ref{eq:def}):
\begin{equation}
\frac{d^3\Gamma(B\to X_f\ell\nu)}{dE_\ell dq^2 dq_0} =\frac{G_F^2|V_{fb}|^2}
{4\pi^3M_B}\,
  \frac{q_0-E_\ell}{\sqrt{|{\bf q}|^2+m_f^2}} \left\{ f(\xi_+)(2\xi_+ 
E_\ell M_B-q^2)
    -(\xi_+ \to \xi_-) \right\}\, ,
\label{eq:triple}
\end{equation}
where
\begin{equation}
\xi_\pm =\frac{q_0\pm\sqrt{|{\bf q}|^2+m_f^2}}{M_B} ,
\label{eq:plumi}
\end{equation}
and $m_f$ is the mass of the decay produced quark, which is the up (charm) 
quark with $f= u (c)$. All lepton masses have been neglected.
Therefore, the distribution function describes leading long-distance QCD 
effects not only in inclusive radiative $B$ decays, but also in inclusive 
semileptonic $B$ decays, and is thus a universal function.  
This universality originates from the fact that the primary object of analysis
in long distance effects is the same bilocal operator matrix element dictated 
by the light-cone dynamics. 

The $b$ quark distribution function $f(\xi)$ defined in 
Eq.~(\ref{eq:def}) differs from the definition given in \cite{jp}.
The difference arises as the distribution function defined here 
is gauge invariant and its inverse Fourier transform is exactly equal to the
leading twist term $F(y^2=0, y\cdot P)$, whereas the distribution function 
of \cite{jp} is defined in the light-cone gauge and its inverse Fourier 
transform is equal to $F(y^2=0, y\cdot P)$ in the leading twist approximation.
However, we note that the expressions for the inclusive semileptonic $B$ decay
rates derived previously \cite{jp} in terms of the distribution function  
remain formally the same, if the distribution function defined in this paper
is used.  This is exemplified by
Eq.~(\ref{eq:triple}) for the triple differential decay rate.

\section{Properties of the distribution function}
The $b$ quark distribution function is an important physical quantity which
summarizes leading long distance effects on inclusive $B$ decay processes.
Equation (\ref{eq:spectrum}) shows that the photon energy spectrum depends
strongly on the distribution function. In Sec.~II we defined the $b$ quark
distribution function in QCD by modifying the definition given in 
Ref.~\cite{jp}. In this section we re-derive several important properties of 
the distribution function. Since it is gauge invariant, it can be calculated  
in any gauge. For simplicity, we
shall choose the light-cone gauge in the following discussion so that the 
Wilson link becomes the identity operator.

It is helpful to introduce the null vector $n^\mu= (1, 0, 0, -1)$ at this 
step, such that the light-like vector $y^\mu= tn^\mu$ with $t$ being a 
parameter. Using this notation, the distribution function defined in 
Eq.~(\ref{eq:def}) can be rewritten as follows
\begin{equation}
f(\xi)= \frac{1}{2\pi}\int dt\, e^{i\xi tn\cdot P}\langle B|b_+^\dagger (0)
b_+(tn)|B\rangle  ,
\label{eq:posi1}
\end{equation}
where $b_+= P_+b$ is the ``good'' component projected out of the $b$ quark 
field, with the projection operator 
$P_+= (1+\gamma^0\gamma^3)/2$.  Inserting a complete set of hadronic states
between quark fields and translating the $tn$ dependence out of the field,
we find
\begin{equation}
f(\xi)= \sum_m\delta[n\cdot(P-\xi P-p_m)]|\langle m|b_+(0)|B\rangle|^2 .
\label{eq:posi2}
\end{equation}
So the distribution function obeys positivity. The state $|m\rangle$ is 
physical and must have $0\leq p_m^0\leq E_B$, thus $f(\xi)= 0$ for $\xi\leq 0$
or $\xi\geq 1$.  Therefore, the support of the distribution function reads
$0\leq\xi\leq 1$. Moreover, one can observe from Eq.~(\ref{eq:posi2}) that
$f(\xi)$ is the probability of finding a $b$ quark with momentum $\xi P$
inside the $B$ meson. This is the familiar probabilistic interpretation of
the parton model \cite{parton} for inclusive $B$ decays.
 
From Eqs.~(\ref{eq:inverse}) and (\ref{eq:decomp}) it is straightforward to 
show that
\begin{equation}
\int_0^1 d\xi\, f(\xi)= F(y=0)= 1 ,
\label{eq:norm}
\end{equation}
because $b$ quantum number conservation implies 
\begin{equation}
\langle B|\bar b\gamma^\mu b|B\rangle = 2P^\mu .
\end{equation}
Thus the distribution
function is exactly normalized to unity, which does not get renormalized as
a consequence of $b$ quantum number conservation.

To understand $B$ meson bound state effects, it is instructive to consider 
first of all the form of the $b$ quark distribution function
in the free quark limit. In this limit, the $B$ meson and the $b$ quark in it
move together with the same velocity: $p_b/m_b= P/M_B\equiv v$ and 
the free Dirac 
field $b(y)= e^{-iy\cdot p_b}b(0)$, so from Eq.~(\ref{eq:def}) it follows that
\begin{equation}
f_{\rm free}(\xi)= \delta(\xi-m_b/M_B) .
\label{eq:delta}
\end{equation}
Substituting Eq.~(\ref{eq:delta}) in Eq.~(\ref{eq:spectrum}), it consistently
reduces to the free quark decay spectrum in the rest frame of the $b$ quark
\begin{equation}
\frac{d\Gamma(b\to s\gamma)}{dE_\gamma}= \frac{G_F^2\alpha m_b^5}{32\pi^4}
|V_{tb}V_{ts}^\ast|^2 |C^{(0)}_7(M_W)|^2\delta (E_\gamma-\frac{m_b}{2}) ,
\label{eq:freesp}
\end{equation}
which is a discrete line at $E_\gamma=m_b/2$.  Reversely, the physical 
photon spectrum is obtained from a convolution of the 
hard perturbative spectrum with the soft nonperturbative distribution function
$f(\xi)$:
\begin{equation}
\frac{d\Gamma (B\to X_s\gamma)}{dE_\gamma}=\int_0^1 d\xi\, f(\xi)
\frac{d\Gamma (b\to s\gamma, p_b=\xi P)}{dE_\gamma} .
\end{equation} 
The bound state smearing of the
photon energy spectrum is then reflected in the deviation of the true 
distribution
function from the delta function form. The quantitative analysis of this
deviation is the subject of the rest of this section.
 
Additional properties of the distribution function can be deduced by means 
of operator product expansion and heavy quark effective theory.
The derivation proceeds in essentially the same manner as in Ref.~\cite{jp}, 
but for the different definition of the distribution function. We present the
corresponding derivation for the new distribution function below for 
completeness.

Since the $b$ quark inside the $B$ meson behaves as almost free due to its 
large mass, relative to which its binding to the light constituents is weak,
one can extract the large space-time dependence
\begin{equation}
b(y)=e^{-im_bv\cdot y}b_v(y) .
\label{eq:scale}
\end{equation}
A Taylor expansion of the field in a gauge-covariant form relates the 
bilocal and local operators. This leads to an operator product expansion
\begin{equation}
\bar b(0)\gamma^\beta b(y)= e^{-im_bv\cdot y}\sum_{n=0}^\infty 
\frac{(-i)^n}{n!}
y_{\mu_1}\cdots y_{\mu_n}\bar b_v(0)\gamma^\beta 
k^{\{\mu_1}\cdots k^{\mu_n\} }b_v(0) ,
\label{eq:moper}
\end{equation}
where $k_\mu=iD_\mu=i(\partial_\mu-ig_sA_\mu)$ and the symbol
$\{\cdots\}$ means symmetrization with respect to the enclosed indices.
Using Lorentz covariance one can express the 
matrix element of the local operator 
on the right-hand side of Eq.~(\ref{eq:moper}) between the $B$ meson states
in terms of the $B$ meson momentum:
\begin{eqnarray}
\lefteqn{\langle B|\bar b_v(0)\gamma^\beta k^{\{\mu_1}\cdots k^{\mu_n\} }
b_v(0)|B\rangle =} \nonumber\\
 & & 2(C_{n0}P^\beta P^{\mu_1}\cdots P^{\mu_n}+
\sum_{i=1}^nM_B^2C_{ni}g^{\beta\mu_i}P^{\mu_1}\cdots P^{\mu_{i-1}}
P^{\mu_{i+1}}\cdots P^{\mu_n}) \nonumber\\
& &+ \ \mbox{terms \ with \ $g^{\mu_i\mu_j}$} .
\label{eq:mlor}
\end{eqnarray}
The terms with $g^{\mu_i\mu_j}$ drop out on the light cone.
Substituting Eqs.~(\ref{eq:moper}) and (\ref{eq:mlor}) in 
Eq.~(\ref{eq:def}) yields
\begin{equation}
f(\xi)=\sum_{n=0}^\infty \frac{(-1)^n}{n!}C_{n0}
\delta^{(n)}(\xi-\frac{m_b}{M_B}) .
\label{eq:mdel}
\end{equation}
Therefore we obtain the moment relation for the distribution function
\begin{equation}
M_n(m_b/M_B)= C_{n0} ,
\label{eq:msum}
\end{equation}
where the $n$th-moment about a point $\tilde{\xi}$ of the  
distribution function is in general defined by
\begin{equation}
M_n(\tilde{\xi})=\int_0^1 d\xi  (\xi-\tilde{\xi})^nf(\xi) .
\end{equation}
By definition, $M_0(\tilde{\xi})=C_{00}=1$. The moment relation (\ref{eq:msum})
relates the moments of the distribution function to the matrix elements of the
local operators in Eq.~(\ref{eq:mlor}).

We invoke heavy quark effective theory to evaluate further expansion 
coefficients in Eq.(\ref{eq:mlor}). 
In this effective theory the QCD $b$-quark field $b(y)$ is related to 
its HQET counterpart $h(y)$ by means of an expansion in
powers of $1/m_b$:
\begin{equation}
b(y)=e^{-im_bv\cdot y}\Bigg [1+\frac{i/ \mkern -12mu D}{2m_b}+
O(\frac{\Lambda^2_{\rm QCD}}{m_b^{2}})\Bigg ]h(y) .
\label{eq:mexp}
\end{equation}
The effective Lagrangian takes the form
\begin{equation}
{\cal L}_{\rm HQET}= \bar hiv\cdot Dh+\bar h\frac{(iD)^2}{2m_b}
h
 +\bar h\frac{g_sG_{\alpha\beta}\sigma^{\alpha\beta}}{4m_b}
h+O(\frac{1}{m_b^2}) ,
\label{eq:mLag}
\end{equation}
where $g_sG^{\alpha\beta}=i[D^\alpha, D^\beta]$ is the gluon field-strength
tensor.
Only the first term in Eq.~(\ref{eq:mLag}) remains in the $m_b\to\infty$
limit, which respects the heavy quark spin-flavor symmetry. The other two 
terms give the $1/m_b$ corrections: the second term violates the heavy flavor
symmetry, while the third term violates both the spin and flavor symmetries.
Using the method described in Refs.~ \cite{chay,bigi,manohar}
to relate matrix elements of local operators in full QCD 
to those in the HQET,
the expansion coefficients $C_{nl}$ in Eq.~(\ref{eq:mlor})
can be expressed in terms of the HQET parameters. 
The nonperturbative QCD effects can,
in principle, be calculated in a systematic manner.
In this formalism the moment $M_n(m_b/M_B)$ is
expected to be of order $(\Lambda_{\rm QCD}/m_b)^n$.
A few coefficients have been evaluated and the results are \cite{jp} 
\begin{eqnarray}
C_{10} & = & \frac{5m_b}{3M_B}E_b +
O(\Lambda_{\rm QCD}^3/m_b^3) , \label{eq:mcoef1} \\
C_{11} & = & -\frac{2m_b}{3M_B}E_b+O(\Lambda_{\rm QCD}^3/m_b^3) , \\
C_{20} & = & \frac{2m_b^2}{3M_B^2}K_b+O(\Lambda_{\rm QCD}^3/m_b^3) , \\
C_{21} & = & C_{22} \ = \  0 , \label{eq:mcoef4}
\end{eqnarray}
where $E_b=K_b+G_b$ and $K_b$ and $G_b$ are the dimensionless HQET parameters 
of order $(\Lambda_{\rm QCD}/m_b)^2$, which are often referred to by the
alternate names $\lambda_1= -2m_b^2K_b$ and $\lambda_2= -2m_b^2G_b/3$,
defined as
\begin{eqnarray}
\lambda_1 &=& \frac{1}{2M_B}\langle B|\bar h
(iD)^2 h|B\rangle , \label{eq:mK} \\
\lambda_2 &=& \frac{1}{12M_B}\langle B|\bar h g_sG_{\alpha\beta}
\sigma^{\alpha\beta} h|B\rangle . 
\label{eq:mG}
\end{eqnarray}
The parameter $\lambda_2$ can be extracted from the $B^\ast - B$ mass 
splitting:
$\lambda_2= (M^2_{B^\ast}-M^2_B)/4=0.12$ GeV$^2$. The parameter $\lambda_1$
suffers from large uncertainty. 

Thus two sum rules for the distribution function can be derived according to 
the moment relation (\ref{eq:msum}). They determine
up to order $(\Lambda_{\rm QCD}/m_b)^2$ the mean value $\mu$ and the variance
$\sigma^2$ of the distribution function, which characterize the location of
the ``center of mass'' of the distribution function and the square of its 
width, respectively:
\begin{equation}
\mu = \frac{m_b}{M_B}\Bigg (1+\frac{5E_b}{3}\Bigg ) ,
\label{eq:mmean}
\end{equation}
\begin{equation}
\sigma^2 = \Bigg (\frac{m_b}{M_B}\Bigg )^2\Bigg [\frac{2K_b}{3}-
\Bigg (\frac{5E_b}{3}\Bigg )^2\Bigg ],
\label{eq:mvar}
\end{equation}
with the definitions
\begin{eqnarray}
& &\mu\equiv M_1(0)=\tilde{\xi}+M_1(\tilde{\xi}) ,
\label{eq:pmean} \\
& &\sigma^2\equiv M_2(\mu)=M_2(\tilde{\xi})-M_1^2(\tilde{\xi}) .
\label{eq:pvar}
\end{eqnarray}
The mean value and variance specify the primary shape of the distribution 
function. Therefore, our evaluation comes to the conclusion that the 
distribution function $f(\xi)$ is 
sharply peaked around $\xi =\mu \approx m_b/M_B$ close to 1 and its width of 
order $\Lambda_{\rm QCD}/M_B$ is narrow.

The sum rules given in Eqs.~(\ref{eq:mmean})
and (\ref{eq:mvar}) quantify the deviation of the true distribution function 
from the delta function form due to bound state effects. 
Choosing the parameters $m_b= 4.9$ GeV and $\lambda_1= -0.5$ GeV$^2$
for the purpose of orientation, we obtain from 
Eqs.~(\ref{eq:mmean}) and (\ref{eq:mvar}) $\mu= 0.933$ and $\sigma^2= 0.006$.
By contrast, the mean value $\mu= m_b/M_B= 0.928$ and the variance
$\sigma^2= 0$ in the free quark limit.

The results for the mean value and the variance given in Eqs.~(\ref{eq:mmean})
and (\ref{eq:mvar}) are at variance with those of \cite{jp} due
to the different definitions of the distribution function.  However, the 
corresponding numerical shifts are found to be so small that the
phenomenological impacts of these differences are negligible. 

Nonperturbative QCD methods such as lattice simulation and QCD sum rules could
help determine further the form of the distribution function.  The distribution
function could also be extracted directly from experiment, as we shall discuss
below.

\section{Measurements of the Distribution Function and Determinations of
the CKM Matrix Elements}
The universality of the distribution function discussed in Sec.~III 
enhances the predictive power of the approach: the
distribution function can be extracted from measurements of one process and
then used to make predictions in all other processes in a model-independent 
manner. In this section we start by exploring how to extract the distribution
function from experiment. Then we will investigate the theoretically clean
methods for the determination of the CKM matrix element $|V_{ts}|$.

The spectrum result, Eq.~(\ref{eq:spectrum}), can be cast in the following 
form:
\begin{equation}
|V_{tb}V^\ast_{ts}|^2 f\Bigg (\frac{2E_\gamma}{M_B}\Bigg ) = 
\frac{2\pi^4 M_B}{G_F^2\alpha m_b^2
|C^{(0)}_7(M_W)|^2}\frac{1}{E^3_\gamma}\frac{d\Gamma(B\to X_s\gamma)}
{dE_\gamma} .
\label{eq:measf1}
\end{equation}
This immediately implies that the distribution function can be extracted 
directly from a 
measurement of the photon energy spectrum upon implementing perturbative 
QCD and electroweak radiative corrections, which have so far been ignored in 
this paper. The experimental data on the photon energy spectrum are already 
available \cite{cleo,aleph,newcleo}, but limited by statistics for a 
meaningful extraction of the distribution function.
Forthcoming very large data samples from high-luminosity $B$ factories
promise a direct extraction of the distribution function with reasonable 
accuracy. 
Such a determination of the distribution function would help substantially to 
reduce the theoretical uncertainties in the
descriptions of both semileptonic and radiative inclusive decays of $B$ 
mesons. The distribution
function extracted from $B\to X_s\gamma$ can be applied, for example, in the
calculations of the lepton energy spectrum and the hadronic invariant mass 
spectrum in charmless inclusive semileptonic decays $B\to X_u\ell\nu$, so
that the precision of the $|V_{ub}|$ determination from these spectra can be 
improved.

The distribution function can also be extracted directly from the measurements
of the differential decay rates as a function of the scaling variable $\xi_+$
[note that $\xi_+$ is different kinematic variable for $B\to X_u\ell\nu$ and
$B\to X_c\ell\nu$, defined in Eq.~(\ref{eq:plumi})]
in the inclusive semileptonic decays of $B$ mesons \cite{jin3}:
\begin{equation}
|V_{fb}|^2 f(\xi_+)= \frac{192\pi^3}{G_F^2M_B^5}\frac{1}
{\xi_+^5\Phi(r_f/\xi_+)}\frac{d\Gamma(B\to X_f\ell\nu)}{d\xi_+} ,
\label{eq:diffxi}
\end{equation}
where $r_f= m_f/M_B \ (f=u,c)$ and $\Phi(x)= 1-8x^2+8x^6-x^8-24x^4{\rm ln}x$.  
Such a determination of the distribution function would also benefit the
theoretical descriptions of both semileptonic and radiative inclusive decays 
of $B$ mesons.  In particular,
once such an extraction of the distribution function is available, it can be
applied as an independent input in turn in the analysis of the 
$B\to X_s\gamma$ photon energy 
spectrum. This would substantially improve the standard model predictions 
and increase the sensitivity to new physics. Therefore, this extraction of the
distribution function is valuable in the search for additional decay
mechanisms beyond the Standard Model in $B\to X_s\gamma$ decays. The 
experimental technique of neutrino reconstruction could well make this way of 
extracting the distribution function experimentally feasible. If the neutrino 
can be reconstructed kinematically by inferring its four-momentum from the 
missing energy and missing momentum in each event, then
it is possible to measure the scaling variable $\xi_+$.

It is important to note that the $B\to X_s\gamma$ photon energy
spectrum and the $B\to X_f\ell\nu$ spectra $d\Gamma/d\xi_+$ share
a common feature. Namely, they offer the intrinsically most sensitive probe 
of long-distance
strong interactions because these spectra correspond to a discrete line 
solely on kinematic grounds in the 
absence of gluon bremsstrahlung and long-distance strong interactions. 
Indeed, our calculation based on the light-cone expansion shows that they are 
explicitly proportional to the 
nonperturbative distribution function. Therefore, the shapes of these spectra
directly reflect the inner long-distance dynamics of the reactions and
measurements of these spectra are idealy suitable for direct extraction of the 
distribution function from experiment. This salient feature also makes the
$b\to u$ scaling variable $\xi_+$ unique to give a very efficient 
discrimination between $B\to X_u\ell\nu$ and $B\to X_c\ell\nu$ 
events \cite{jin3}, even better than the hadronic invariant mass.

Given the difficulty in distinguishing $t\to s$ decays from the
dominant $t\to b$ decay mode, it is hard to determine $|V_{ts}|$ from studies 
of top quark decays. Based on the results obtained in this paper and in 
Ref.~\cite{jin3}, we propose a new strategy to extract $|V_{ts}|$ which is 
largely free of hadronic uncertainties.
The idea is to use the known normalization of the distribution function
or the cancellation of the distribution function in the ratio of the decay
rates to eliminate the dominant hadronic uncertainties.  
The most straightforward and best way to eliminate the dependence on the 
distribution function is to resort to, again, the $B\to X_s\gamma$ 
photon energy spectrum and the $B\to X_{u,c}\ell\nu$ spectra $d\Gamma/d\xi_+$
since they are proportional to the distribution function.  

Integrating Eq.~(\ref{eq:measf1}) over 
$E_\gamma$ and using the normalization condition (\ref{eq:norm}) yields
\begin{equation}
|V_{tb}V_{ts}^\ast|^2 = \frac{4\pi^4}{G_F^2 \alpha m_b^2 
|C^{(0)}_7(M_W)|^2}
\int_0^{M_B/2} dE_\gamma \frac{1}{E_\gamma^3}\frac{d\Gamma(B\to X_s\gamma)}
{dE_\gamma} .
\label{eq:vts}
\end{equation}
Thus the known normalization of the distribution function allows an almost 
model
independent determination of $|V_{tb}V_{ts}^\ast|$ from a measurement of the
weighted integral of the photon energy spectrum. 
This determination can be taken as a measurement of
$|V_{ts}|$ by using $|V_{tb}|= 0.9993$ from unitarity of the CKM matrix,
which holds to a very high accuracy \cite{PDG}.  The advantage of this 
determination of $|V_{ts}|$ is that the dominant hadronic uncertainty has been
avoided, which may provide one of the most precise determinations of 
$|V_{ts}|$. 

As in the case 
of $B\to X_s\gamma$ discussed above, one can get rid of the distribution
function by integrating Eq.~(\ref{eq:diffxi}) over $\xi_+$ \cite{jin3}:
\begin{equation}
|V_{fb}|^2= \frac{192\pi^3}{G_F^2M_B^5}\int_{r_f}^1 d\xi_+ \frac{1}
{\xi_+^5\Phi(r_f/\xi_+)}\frac{d\Gamma(B\to X_f\ell\nu)}{d\xi_+} .
\label{eq:vqb}
\end{equation}
Thus, the precise determinations of $|V_{ub}|$ and $|V_{cb}|$ can be 
obtained from the measurements of the weighted integrals of the differential 
decay rates as functions of $\xi_+$ in $B\to X_u\ell\nu$ and $B\to X_c\ell\nu$,
respectively.

Alternatively, when we take the ratio of the differential decay 
rates, the distribution function cancels. From Eqs.~(\ref{eq:measf1}) and
(\ref{eq:diffxi}), we obtain
\begin{equation}
\left|\frac{V_{tb}V_{ts}^\ast}{V_{fb}}\right|^2= 
\frac{\pi M_B}{3\alpha m_b^2|C^{(0)}_7(M_W)|^2}E^2_\gamma\Phi(r_f/\xi_+)
\frac{d\Gamma(B\to X_s\gamma)/dE_\gamma}
{d\Gamma(B\to X_f\ell\nu)/d\xi_+}\Bigg |_{E_\gamma=M_B\xi_+/2} ,
\end{equation}
which may be useful to provide a theoretically clean determination of 
$|V_{ts}/V_{fb}|$. By the same token, an analogous expression has 
been derived 
in \cite{jin3} for $B\to X_{u,c}\ell\nu$, which can be used to measure 
$|V_{ub}/V_{cb}|$ to a high precision. These determinations of the CKM matrix
elements rely on the universality of the distribution function, in contrast 
to the method described above by virtue of the known normalization of the
distribution function. The compatibility between the
various measurements will test the universality of the distribution function 
in the inclusive
semileptonic and radiative decays of $B$ mesons within the Standard Model.

\section{Summary}
In this paper we have calculated the long distance effects on the photon 
energy spectrum
in $B\to X_s\gamma$ decays. We have demonstrated on the basis of light-cone 
expansion that the leading nonperturbative QCD contribution in inclusive 
radiative $B$ decays with emission of a sufficiently high energy photon 
resides 
in the distribution function. The distribution function is defined by Fourier
transformation of the matrix element of the non-local $b$ quark operators
separated along the light cone. We have found that the distribution function 
is universal in the sense that the same distribution function also summarizes 
the leading nonperturbative QCD contribution in 
inclusive semileptonic $B$ decays at large momentum transfer. 

Although long-distance strong interactions responsible for the distribution
function preclude a complete calculation of it at present, we have deduced
some of its basic properties in QCD.
The distribution function is gauge invariant and obeys positivity. It has a 
support between 0 and 1 and is
exactly normalized to unity because of $b$ quantum number conservation.
It contains the free quark decay as a limiting case with 
$f_{\rm free}(\xi)= \delta
(\xi-m_b/M_B)$. The distribution function $f(\xi)$ can be interpreted as the 
probability of finding a $b$ quark with momentum $\xi P$ inside the $B$ meson.
In addition, we evaluated
the mean and variance of the distribution function using the
techniques of operator product expansion and heavy quark effective theory.
They specify the primary shape of the distribution function and quantify
the deviation from the delta function form in the free quark limit.

The $b$ quark distribution function for the $B$ meson is a key object. Like
the well-known parton distribution functions for the nucleon in deeply
inelastic lepton-nucleon scattering, the knowledge of it would help us 
greatly in understanding the nature of confinement and the structure of the $B$
meson. One should
try to calculate it using nonperturbative QCD methods such as lattice
simulation and QCD sum rules. On the other hand, the distribution function
can be extracted directly from experimental data. 
The underlying common structure of 
long-distance strong interactions correlates the $B\to X_s\gamma$ and 
$B\to X_f\ell\nu$ processes.  
The universality of the distribution function implies that once it is
measured from one process, it can be used to make predictions in all 
other processes in a model-independent manner. 
We have discussed the direct extraction of the distribution
function from the measurements of the $B\to X_s\gamma$ photon spectrum or
the $B\to X_f\ell\nu$ spectra $d\Gamma/d\xi_+$.  We stress that these decay
spectra are unique in that they offer the intrinsically most sensitive probe
of long-distance strong interactions. 
The experimental extraction of the distribution function will lead to a 
significant improvement of the theoretical description of the 
$B\to X_s\gamma$ photon energy spectrum, which is very important for seeking
new physics. The extracted distribution function will also
improve the theoretical description of inclusive semileptonic $B$ decays,
allowing for more precise determinations of $|V_{ub}|$ and $|V_{cb}|$.
Moreover, a confrontation of experimental extraction of the 
distribution function with QCD predictions will be a test of our understanding
of the $B$ meson structure and nonperturbative techniques.
A direct extraction of the distribution function will therefore
be an important aspect in future experiments in inclusive $B$ meson decays.

Measurements of $B\to X_s\gamma$ decays can be used to determine the CKM 
matrix element $|V_{ts}|$. We have described the theoretically clean methods
for the determinations of the CKM matrix elements by avoiding the dominant 
hadronic
uncertainties using the known normalization of the distribution function or
the cancellation of the distribution function in the ratio of the differential
decay rates. The residual hadronic uncertainty in $|V_{ts}|$ due to 
higher-twist, power-suppressed corrections is expected to be at the level of
one percent. With hadronic uncertainties well under control, these methods 
eventually will yield the most accurate value of $|V_{ts}|$, which is, on the
other hand, probably
one of the most reliable quantities to signal new physics in $B\to X_s\gamma$.
The agreement of $|V_{ts}|$ extracted from 
the flavour changing neutral current process $B\to X_s\gamma$ with the value
obtained from the direct measurements plus unitarity under the assumption that
the Standard Model is valid can be used to impose constraints on new physics 
models. 

The universality of the distribution function in radiative and
semileptonic inclusive decays of $B$ mesons is valid only to the leading order 
in the light-cone expansion. To what extent this is a good approximation 
can be tested experimentally in a variety of ways
in the $B\to X_s\gamma$, $B\to X_u\ell\nu$ and $B\to X_c\ell\nu$
processes. Higher twist effects may be numerically sizable in some region of 
phase space, especially in the resonance domain.
Their quantitative consequences deserve a thorough investigation.

The calculation of long distance effects presented in this paper must be 
combined with
the perturbative QCD and electroweak radiative corrections to give a detailed 
theoretical description of the photon energy spectrum. 
The theoretical development toward a treatment of inclusive radiative decays 
of $B$ mesons from first principles, in conjunction with precision measurements
made possible by $B$ factories in the coming few years, will make
a decisive test of the Standard Model at one-loop level and, more
excitingly, might corroborate the existence of new physics in inclusive 
radiative decays of $B$ mesons.

\acknowledgments
Useful discussions with Xiao-Gang He and Emmanuel Paschos are gratefully
acknowledged. This work was supported by the Australian Research Council. 

{\tighten

} 

\end{document}